\documentclass[preprint2]{aastex62}

\usepackage{graphicx}
\usepackage[suffix=]{epstopdf}
\usepackage{natbib}
\usepackage{amsmath}
\usepackage{url}
\usepackage{xspace}

\newcommand{\Kepler}{\textsl{Kepler}\xspace}

\begin{document}
\title{Rotating Stars from \Kepler Observed with Gaia DR2}

\shorttitle{Rotating Stars from \Kepler Observed with Gaia DR2}
\shortauthors{Davenport \& Covey}

\correspondingauthor{James. R. A. Davenport}
\email{James.Davenport@wwu.edu}

\author{James. R. A. Davenport}
\altaffiliation{NSF Astronomy and Astrophysics Postdoctoral Fellow}
\altaffiliation{DIRAC Fellow}
\affiliation{Department of Physics \& Astronomy, Western Washington University, 516 High St., Bellingham, WA 98225, USA}
\affiliation{Department of Astronomy, University of Washington, Seattle, WA 98195, USA}

\author{Kevin R. Covey}
\affiliation{Department of Physics \& Astronomy, Western Washington University, 516 High St., Bellingham, WA 98225, USA}

\begin{abstract}
We have matched the astrometric data from {\em Gaia} Data Release 2 to the sample of stars with measured rotation periods from \Kepler. Using 30,305 stars with good distance estimates, we select 16,248 as being likely main sequence single stars centered within a 0.5 mag region about a 1 Gyr isochrone, removing many sub-giants and unresolved binary stars from the sample.
The rotation period bimodality, originally discovered by \citet{mcquillan2013}, is clearly recovered for stars out to 525pc, but is not detectable at further distances. This bimodality correlates with Galactic height as well, dropping strongly for stars above $Z>90$ pc.
We also find a significant width in the stellar main sequence of $\Delta M_G\sim$0.25 mag, as well as a coherent gradient of increasing rotation periods orthogonal to the main sequence.
We interpret this as a signature of stellar angular momentum loss over time, implying a corresponding diagonal age gradient across the main sequence. Stellar evolution models predict changes in color and luminosity that are consistent in amplitude, but not in direction, with those required to produce the gradient we have detected. This rotation gradient suggests that main sequence evolution produces offsets in color--magnitude space that are significantly more orthogonal to the zero-age main sequence than models currently predict, and may provide new tests for both stellar evolution and gyrochronology models.
\end{abstract}

\section{Introduction}
Over the past decade the study of stellar rotation has transformed from a niche observation limited to a handful of nearby clusters and few thousand bright stars, to an area of great  interest for field stars with a rapidly growing sample size.
The \Kepler mission \citep{borucki2010}, through its unmatched combination of photometric precision and years-long observation baseline, has produced the largest precision sample of rotating stars to date -- more than 30,000 sources to date \citep{mcquillan2014}. 
Although the light curves available from the {\em Gaia} mission are not as densely sampled in time as \Kepler \citep{gaia}, they have yielded over 147,000 candidate rotation periods as of Data Release 2 \citep{lanzafame2018}. 
The extended \Kepler mission, K2 \citep{howell2014}, has now observed more stars than the original \Kepler sample, potentially doubling the number of high precision stellar rotation periods available. 
Similarly, by the end of the {\em Gaia} mission \citet{lanzafame2018} estimate a sample of 3--20 million rotation periods will be recovered. The era of statistical studies in stellar rotation has therefore begun.

Stellar rotation has long been noted as a means to possibly age-date stars due to their constant angular momentum loss via winds \citep{skumanich1972}. 
While studies of open clusters give hope that this ``gyrochronology'' model broadly works for solar-type and lower-mass stars, many uncertainties exist about the details of this spin-down and its utility as a clock. These include questions about the initial rotation period distribution for stars \citep[e.g.][]{barnes2010,matt2015}, the specific analytic prescription for modeling the spin-down \citep{angus2015}, and exploring the efficiency of this angular momentum loss mechanism at older ages \citep{van-saders2016}.

One of the most compelling results from the rotating star sample in \Kepler is the discovery of a bimodal period distribution.
\citet{mcquillan2013} first found a bimodality in the distribution of field M dwarfs with periods between $\sim$10 and $\sim$50 days. This feature was also found in the \Kepler field K dwarfs in \citet{mcquillan2014}, but was not seen in the bluer stars. Using Gaia DR1, \citet{davenport2017} was able to remove contaminating sub-giants from the rotation sample, and found the bimodality extended to the G dwarfs as well. This bimodal surface rotation period distribution is either a new short-lived transition or instability phase of rapid angular momentum loss, or a signature of star formation history imprinted in the present-day rotation period distribution. 

However, as \citet{davenport2017} notes this feature has only been observed in the \Kepler rotation period catalog, and most critically only for stars within $\sim$300 pc. The faint M dwarfs in the \Kepler sample studied in \citet{mcquillan2013} have an average distance of $<$250 pc. Gaia DR1 was also only able to provide precise distances for the nearest stars, and \citet{davenport2017} report an median distance of 285 pc for the blue stars in their sample.

In this paper we extend the work of \citet{davenport2017} in studying the \Kepler rotation period distribution as a function of distance from the Sun using new astrometric data from the {\em Gaia} mission. 
By matching the \citet{mcquillan2014} rotation period catalog to the newest data from {\em Gaia} Data Release 2 \citep{gaia_dr2}, we can use precise distances for essentially every star in the McQuillan catalog to select the most-likely main sequence dwarfs out to distances $>$2 kpc. Importantly this filters out both sub-giants, the main contaminant noted by \cite{davenport2017}, and unresolved binary stars.
Here we demonstrate the power of such a combined time-domain and astrometric sample for constraining the detailed evolution of main sequence stars themselves, and exploring the star formation history of the Milky Way.

\section{The \Kepler--Gaia Data}

We used the largest homogeneous catalog of rotation periods available from the \Kepler mission. The sample from \citet{mcquillan2014} provides rotation periods for more 34,030 stars, measured using the Auto-Correlation Function (ACF). While the ACF does not recover periods with as much precision as methods such as the Lomb-Scargle Periodogram, it is more robust to detecting the true period as opposed to an alias, and more complete for batch analysis of all stars \citep[e.g. see][]{aigrain2015}.

The \Kepler data was matched to the {\em Gaia} DR2 source catalog using a 1 arcsecond radius. We used the \Kepler--Gaia cross-match made publicly available by M. Bedell, which included entries for 195,830 sources. \Kepler-based stellar parameters are included in this cross-match from the Data Release 25 \Kepler catalog. Joining this cross-matched table to the \citet{mcquillan2014} catalog, we found 33,538 sources with {\em Gaia} astrometry and \Kepler-derived rotation periods.

To select stars with good parallaxes, as well as high quality photometry from {\em Gaia}, we selected stars with the following criteria:
\begin{itemize}
\item Parallax error $< 0.1$ mas
\item $\sigma(M_{G}) / M_{G} < 0.01$
\item $\sigma(G_{BP}) /G_{BP} < 0.01$
\item $\sigma(G_{RP}) /G_{RP} < 0.01$
\end{itemize}

Rather than simply use the inverse {\em Gaia} parallax values to measure the distance to sources, we use the improved distance prescription from \citet{bailer-jones2018}, who provided independent distance estimates for 1.33 billion {\em Gaia} sources using a weak prior on the distribution of stars in our Galaxy. We follow their suggested use of the distance catalog, including only sources with {\tt modality\_flag} == 1 (i.e. not a bimodal distance solution) and {\tt result\_flag} == 1 (i.e. a well constrained distance).

\begin{figure}[]
\centering
\includegraphics[width=3.6in]{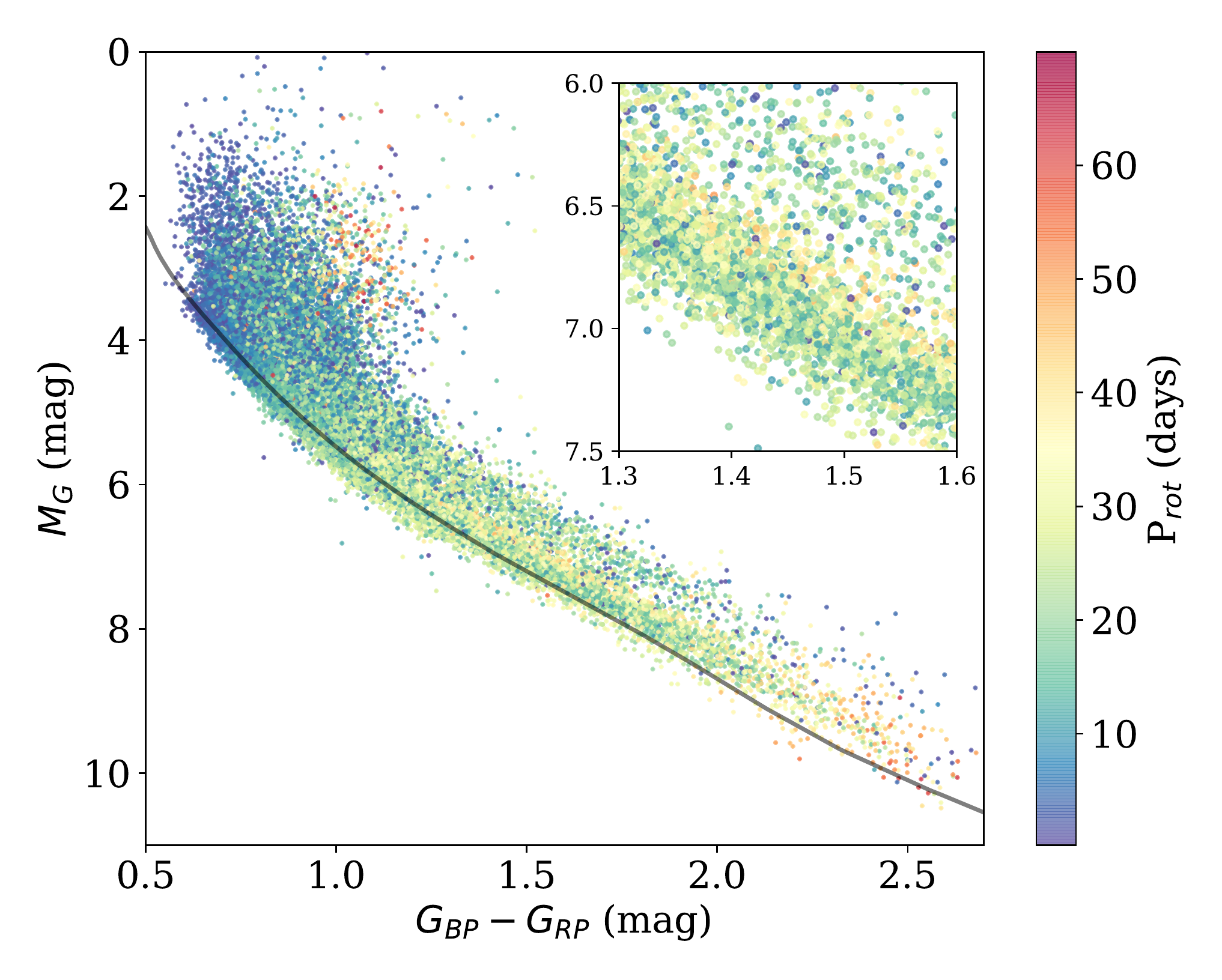}
\caption{
Color--magnitude diagram for 30,305 \Kepler stars from the \citet{mcquillan2014} sample that are included in {\em Gaia} DR2, colored by their measured rotation period. For reference we show a $10^9$ year MIST isochrone used to select likely main sequence, single stars (black line). As in \citet{davenport2017}, the bluer stars in the sample are significantly contaminated by sub-giants. 
A track of binary stars is apparent $\sim$0.75 mag above the main sequence. The inset figure highlights a portion of the single star and equal-mass binary main sequences.
}
\label{fig:cmd}
\end{figure}

Our final sample contained 30,305 stars in {\em Gaia} DR2 with measured \Kepler rotation periods that passed these selection criteria. A color--magnitude diagram (CMD) of this sample using the {\em Gaia} bands is presented in Figure \ref{fig:cmd}, with points colored by their measured \Kepler rotation periods.

\section{Selecting Main Sequence Stars}

As in \citet{davenport2017}, the color--magnitude diagram in Figure \ref{fig:cmd} shows many of the bluer stars in the \citet{mcquillan2014} sample are located significantly above the main sequence. These are likely subgiant stars, which no longer follow the main sequence stars spin-down evolution \citep[e.g.][]{donascimento2012, van-saders2013}. Since \citet{davenport2017} found subgiants could obscure the rotation period bimodality for G dwarfs, these must be excluded from our analysis, but we encourage future studies to explore the wealth of angular momentum evolution data from these post-main sequence objects.

Beyond the subgiant contamination, we also see a secondary population of stars in a parallel track $\sim$0.75 mag above the normal main sequence, as expected and seen in other color--magnitude diagrams. This parallel main sequence occurs due to unresolved equal-mass (or nearly equal-mass) field binaries, and was seen in the {\em Gaia} DR1 data as well \citep{anderson2017}. Since the tightest of these systems may have experienced tidal evolution that could significantly impact their rotation evolution \citep[e.g.][]{lurie2017}, and we lack the ability with the {\em Gaia} DR2 data to adequately constrain their physical separations,
we must also remove these unresolved binaries from our analysis. We do not explore the binary population in any detail here, but this sample could provide useful insight into the tidal evolution of binary stars, and are good targets for radial velocity follow-up to characterize binary system properties. We also note a small number of systems above the even the equal-mass binary main sequence track, which could be due to unresolved triple star systems.

We use an isochrone from the Mesa Isochrones and Stellar Tracks suite \citep[MIST;][]{MIST} to choose likely main sequence stars in Figure \ref{fig:cmd}. Our favored model to represent the main sequence in this study had [Fe/H] = +0.25 and an age of $10^9$ years, and was chosen by-hand. Single, main sequence stars were selected in a region spanning 0.1 mag fainter and 0.4 mag brighter than the MIST isochrone, resulting in a final sample of 16,248 stars for analysis of their rotation period distributions.

\section{Tracing the Period Bimodality}

Using this sample of likely single, main sequence stars from \Kepler and {\em Gaia}, we are able to explore the distribution of rotation periods for stars as a function of their distance and Galactic location. The rotation period bimodality in \Kepler stars was previously detected only for stars within $\sim$300 pc of the Sun due to the limits of available parallax data. Now with {\em Gaia} DR2, our sample of main sequence stars with measured rotation periods from \Kepler with adequate distance estimates for filtering out sub-giants extends to over 2 kpc.

\begin{figure*}[]
\centering
\includegraphics[width=3.5in]{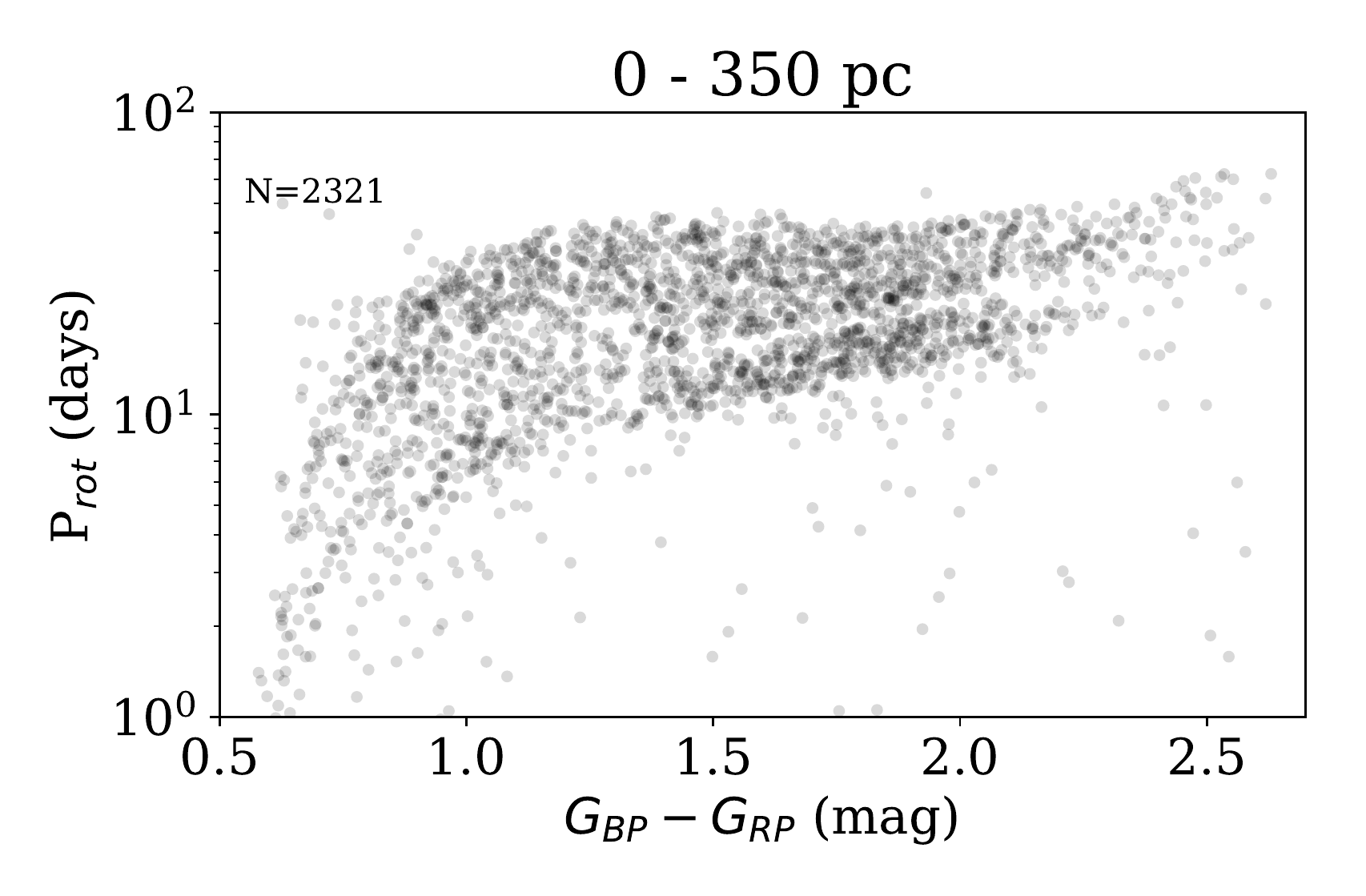}
\includegraphics[width=3.5in]{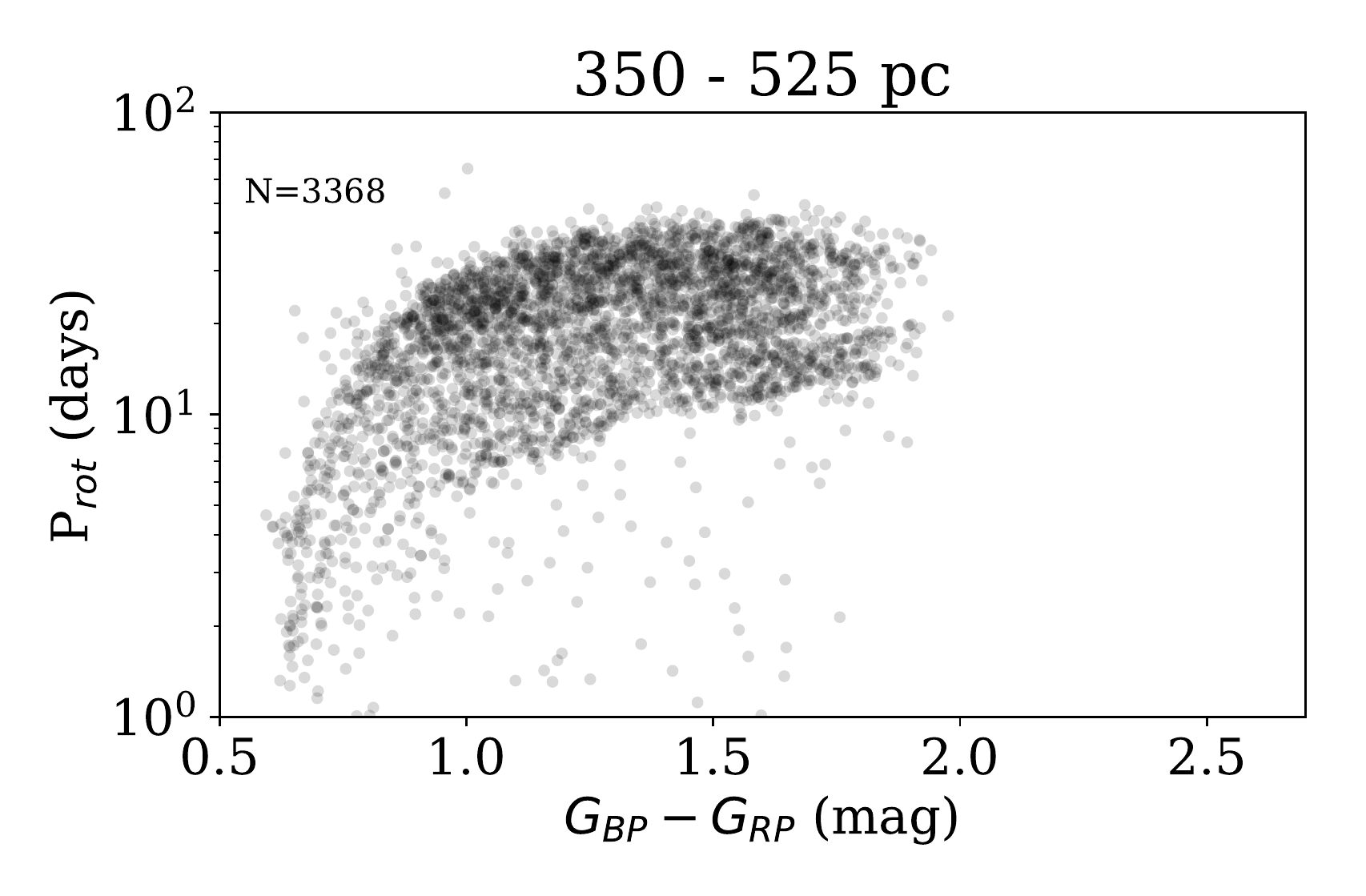}
\includegraphics[width=3.5in]{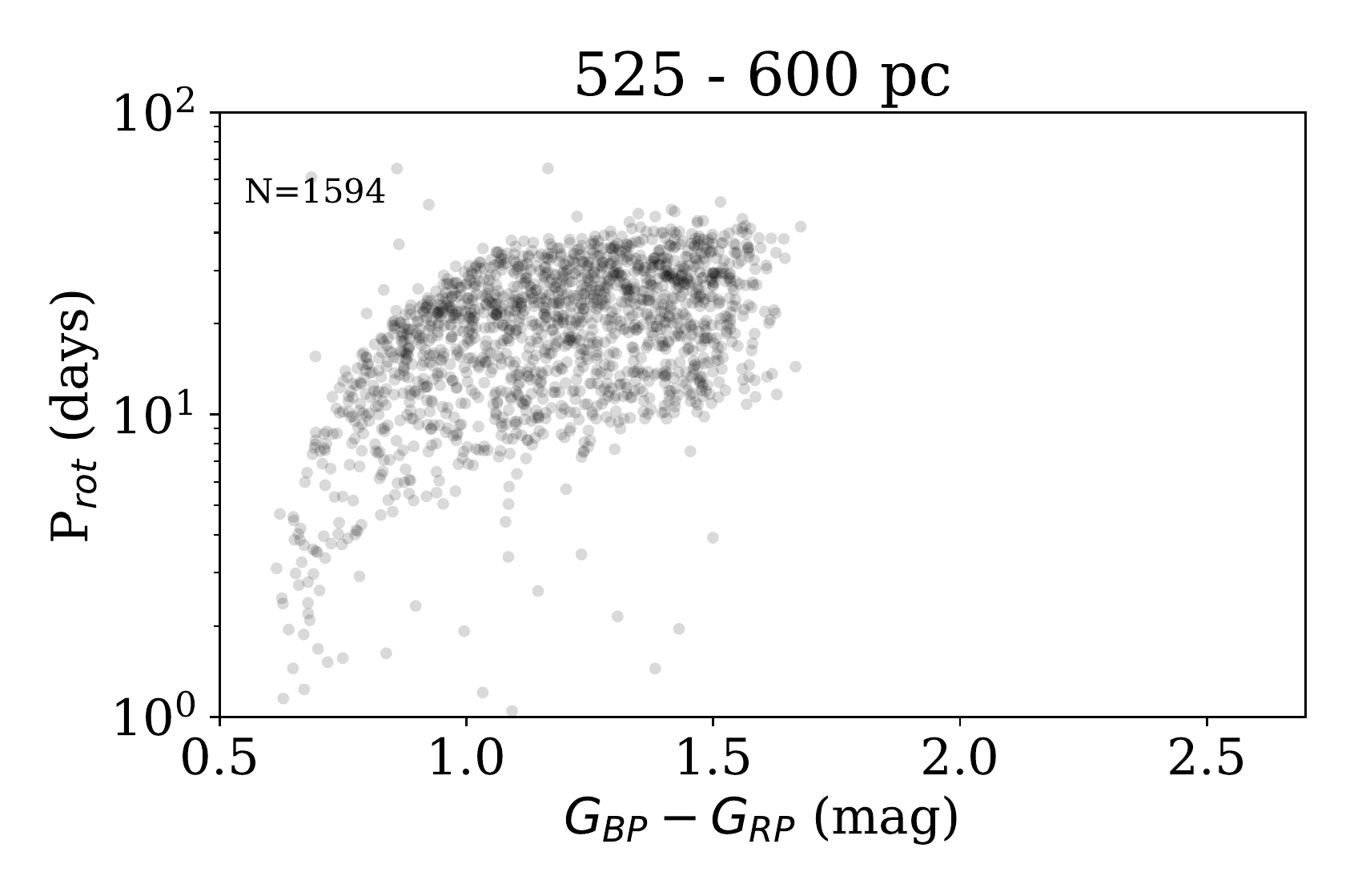}
\includegraphics[width=3.5in]{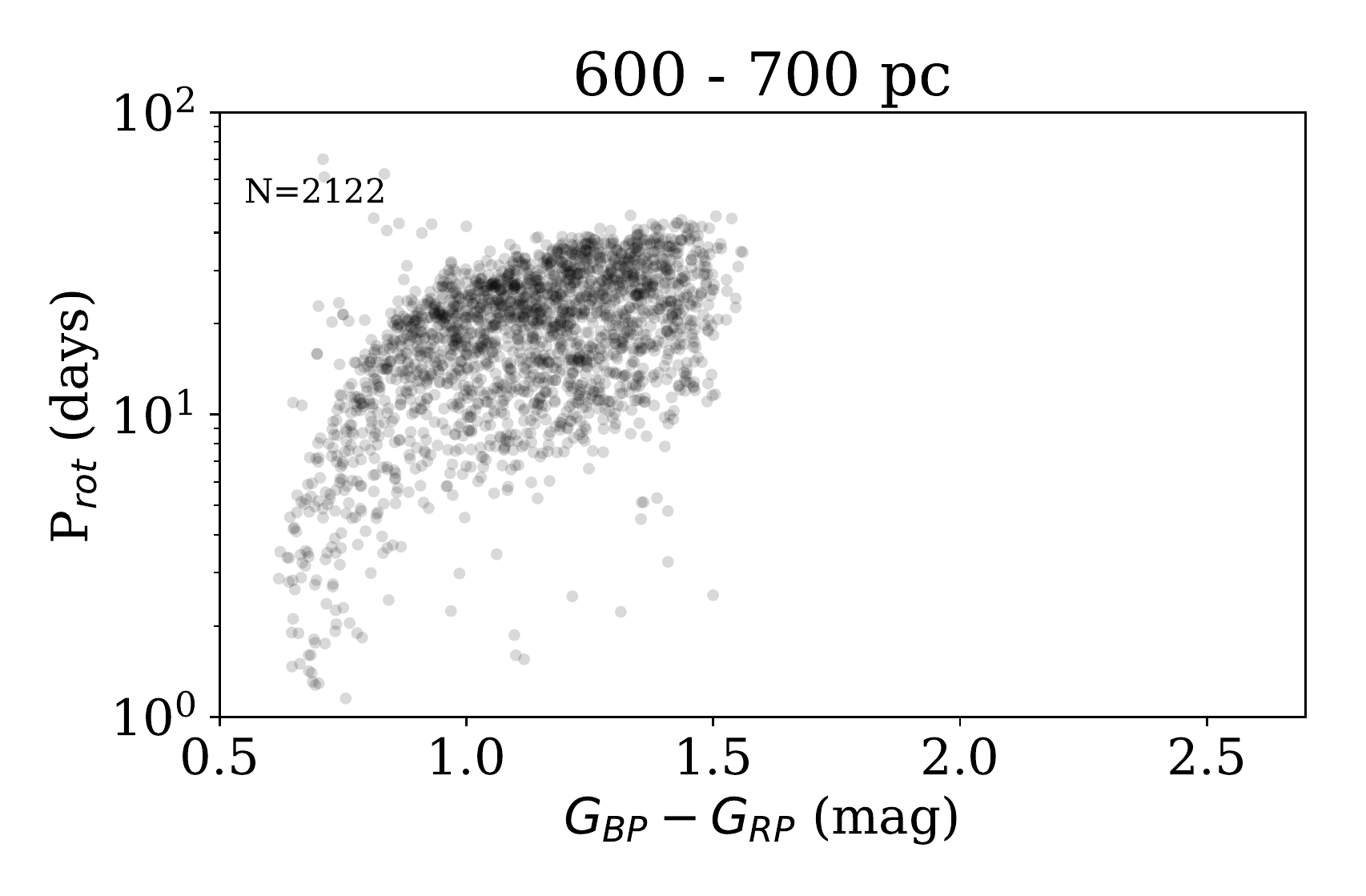}
\includegraphics[width=3.5in]{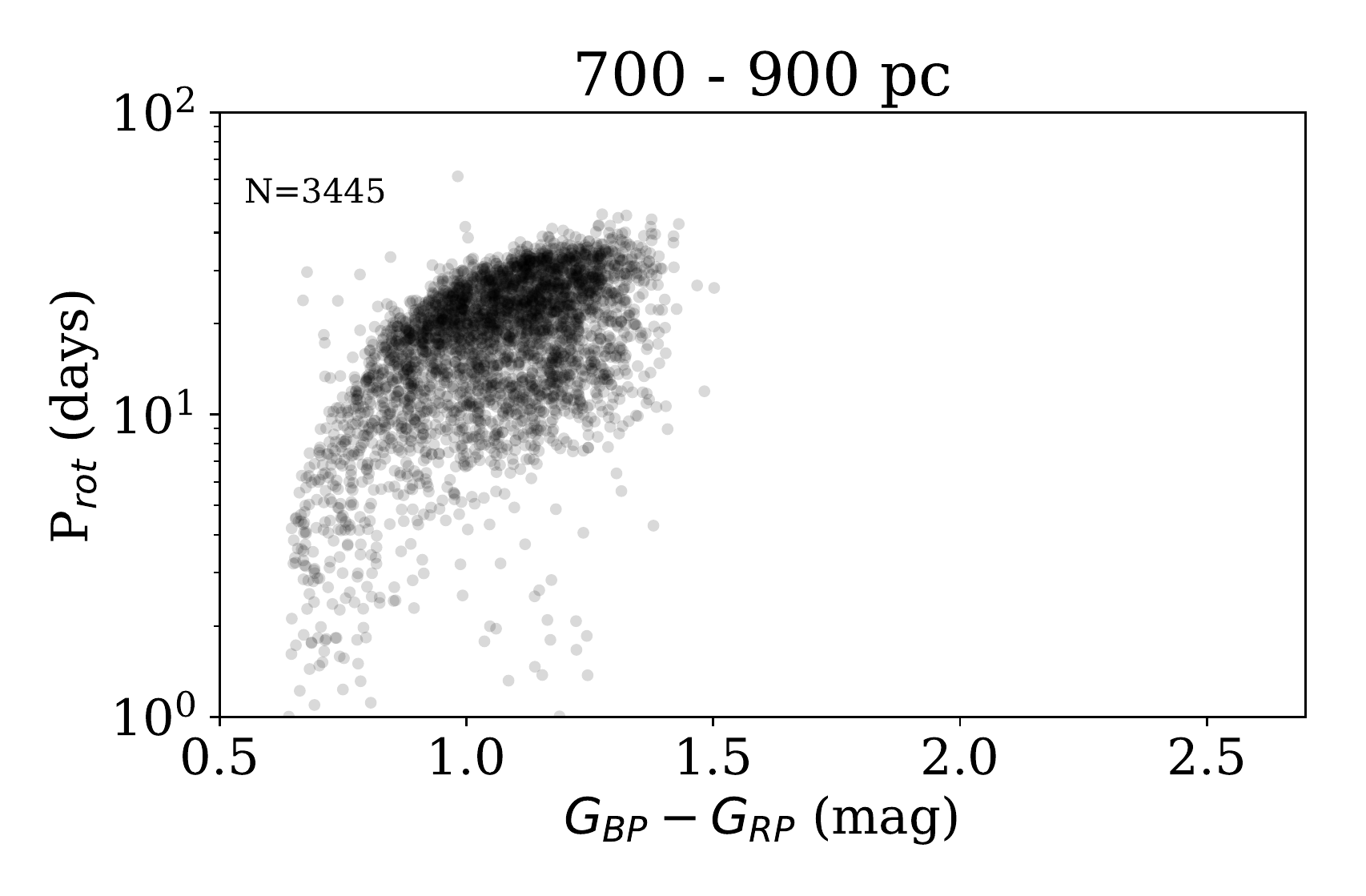}
\includegraphics[width=3.5in]{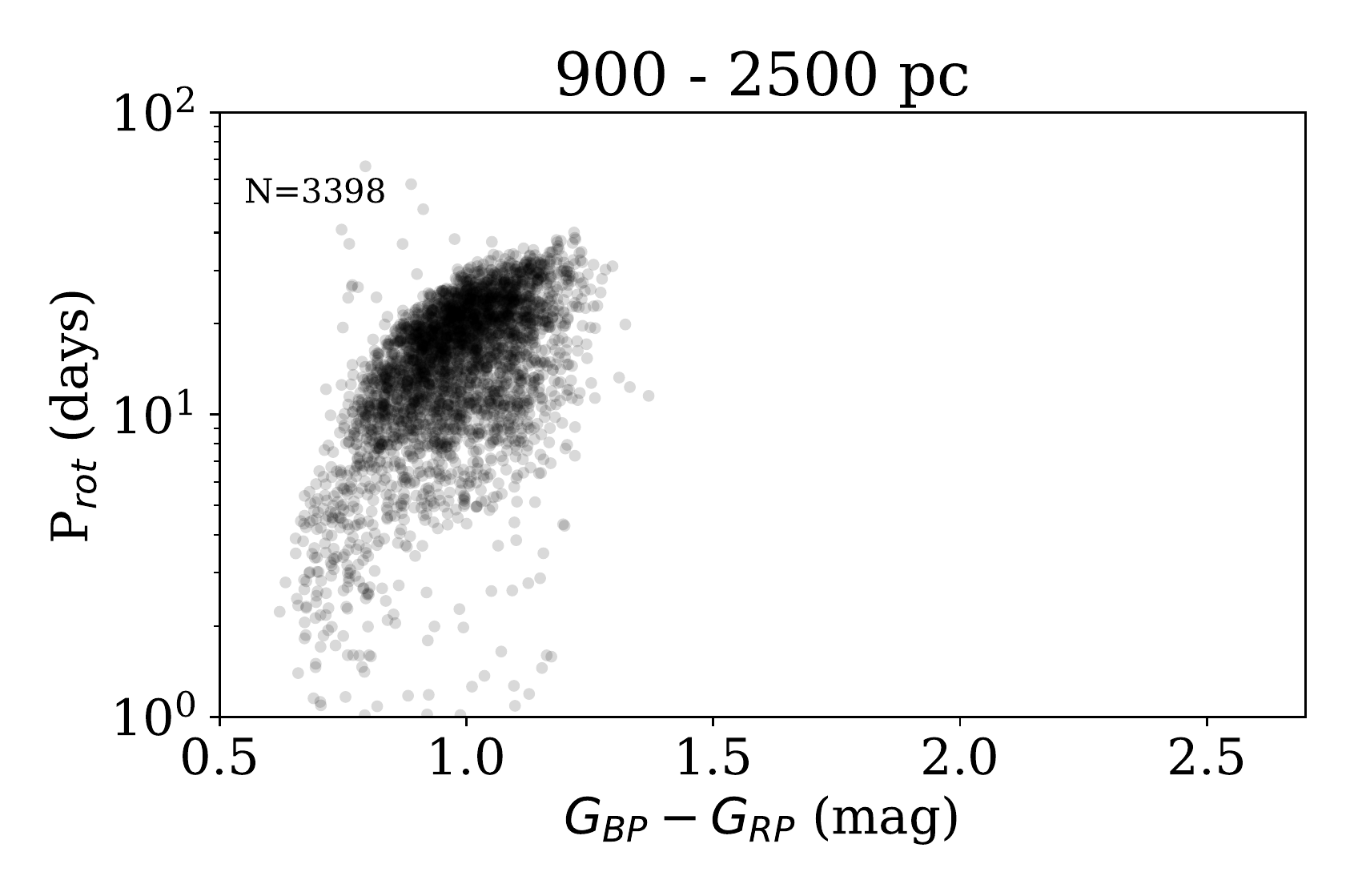}
\caption{Color--period diagrams for our sample of likely main sequence stars, divided into six bins of distance. Our nearest bin (within 350pc) is effectively the distance analyzed in \citet{davenport2017} using Gaia DR1, and clearly shows the rotation period bimodality for the entire sample. The brighter magnitude limit of the \Kepler sample results in redder (fainter) stars missing in our further distance bins. The rotation period bimodality can be seen in the 350-525 pc bin, but is not found in the bluer stars at further distances.
}
\label{fig:color_period}
\end{figure*}

In Figure \ref{fig:color_period} we present the period--color diagram for our sample of stars, split into six bins of projected distance. The first panel (0--350 pc) effectively reproduces the results of \citet{davenport2017} for bluer stars and \citet{mcquillan2014} for the redder stars. A gap in the observed rotation periods as a function of color is seen, at a period of approximately 5 days for $G_{BP}-G_{RP}\approx1$, 20 days for $G_{BP}-G_{RP}\approx2$, and increasing towards 30 days for the reddest stars in our sample. This gap corresponds with a line of approximately constant age, consistent with a gyrochrone with age $\sim$600 Myr \citep{davenport2017}.

\begin{figure*}[!ht]
\centering
\includegraphics[width=3.5in]{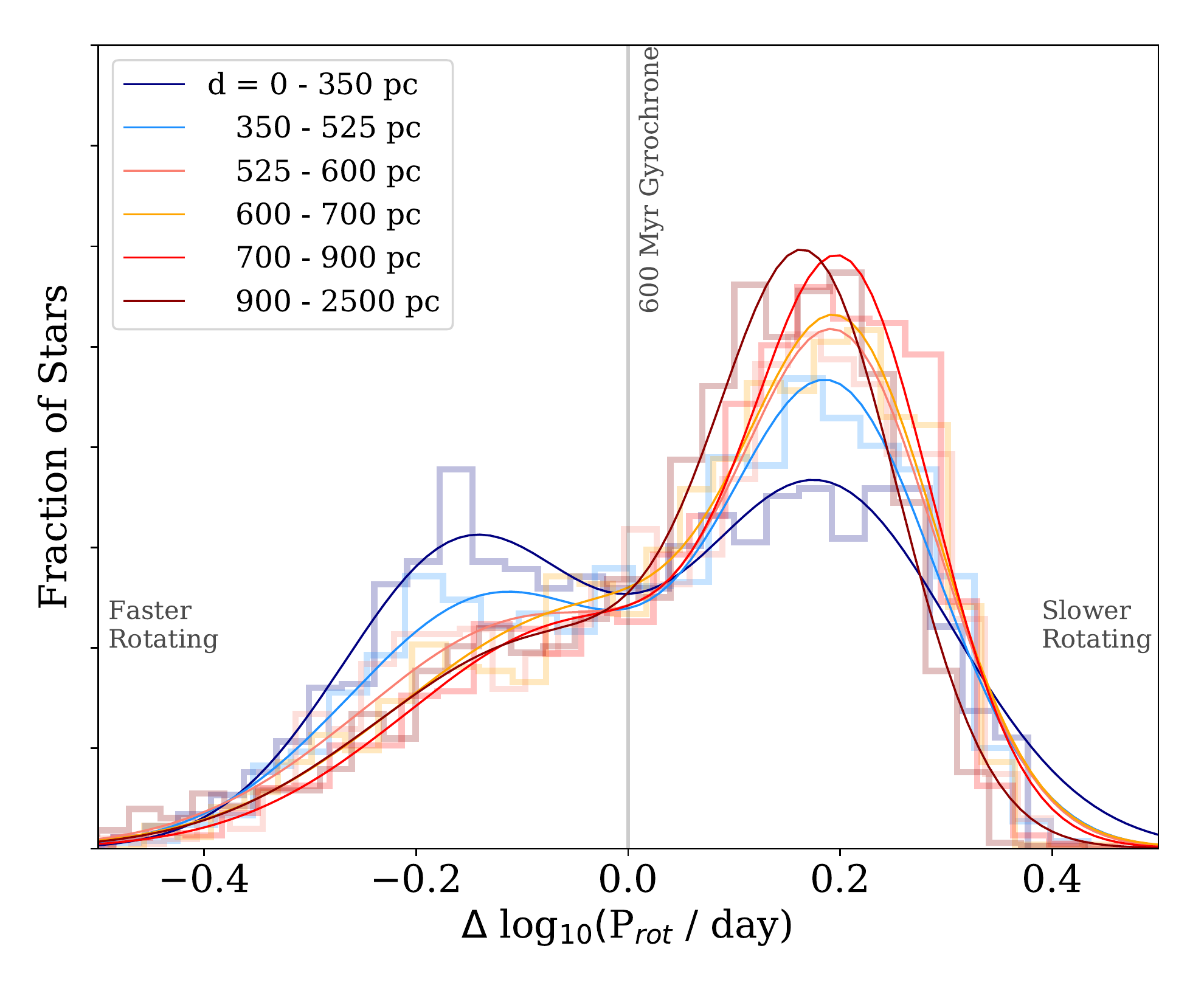}
\includegraphics[width=3.5in]{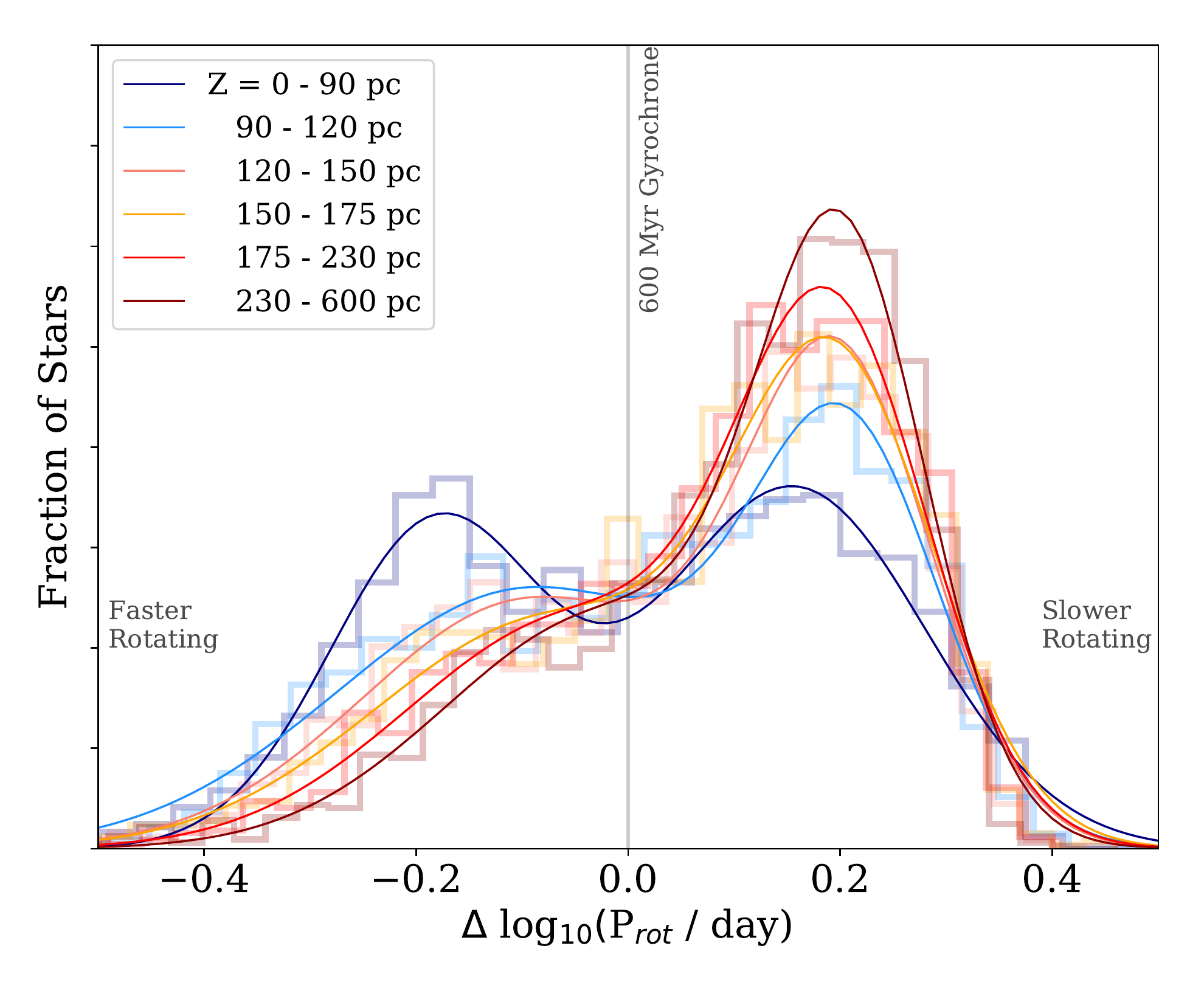}
\caption{Left: Distribution of the log rotation periods after a 600 Myr gyrochrone was subtracted, using the same distance bins shown in Figure \ref{fig:color_period} (faint lines). Two-gaussian models were fit to each histogram (bold lines). 
The rotation period bimodality for stars within 350 pc has two nearly equal peaks similar to those found in \citet{davenport2017}, at -0.15 and +0.18 dex. The fast rotating peak (left side) declines sharply at further distances.
Right: Same as Left, but in bins of Galactic height above the plane (Z). The drop-off of fast rotating, young stars is even more pronounced as a function of height.
}
\label{fig:per_hist}
\end{figure*}

The bimodality is still apparent in the second distance bin (350--525 pc), clearly visible in the redder stars, but seems to fade in the final three bins. Other structures in the period distributions are visible, however. For example, a thin sequence of stars with rotation periods near 10 days is faintly visible in the most distant bin (900--2500 pc) for stars with colors of $0.8<G_{BP}-G_{RP}<1.2$. This feature is due to the 1 Gyr open cluster NGC 6811 in the \Kepler field \citep{meibom2011}, whose distance is $\sim$1100 pc \citep{sandquist2016}.

To better illustrate the evolution of rotation periods for all the stars between the distance bins, in Figure \ref{fig:per_hist} we follow \citet{davenport2017} and subtract the rotation period of a 600 Myr gyrochrone. As no published gyrochronology model yet exists that has been tuned to the {\em Gaia} photometric colors, we adopt the same gyrochronology model of Eqn. 2 from \citet{meibom2009} used by \citet{davenport2017} to approximately trace the rotation period gap at 600 Myr as a function of $B-V$ color. We convert stars from the observed {\em Gaia} $G_{BP}-G_{RP}$ color to $B-V$ using the same 1 Gyr MIST isochrone model used to define the main sequence in Figure \ref{fig:cmd} above. 

Our sample naturally becomes biased towards the bluer (brighter) stars as we reach larger distances. Since bluer stars have more rapid rotation and more dramatic period evolution, gyrochronology models are ``bent'' strongly for these stars, and thus the period bimodality (or other structures) become difficult to distinguish. \citet{davenport2017} used an ad-hoc modification of the gyrochrone model for the hottest stars to illustrate the existence of the period bimodality. However, in Figure \ref{fig:per_hist} we limit our analysis to stars with $0.8<B-V<1.5$ (approximately $0.9<G_{BP}-G_{RP}<2.2$, or $0.5<M_\odot< 0.9$), where the gyrochrone is most flat as a function of color. This color range was chosen to ensure ample stars were available in each distance bin, but without having to use the ad-hoc gyrochrone model correction from \citet{davenport2017}.

Figure \ref{fig:per_hist} (left) shows clearly that the short period component of the rotation period bimodality dominates only for the nearest stars, and steadily decreases with increasing distance.
The \Kepler field is centered on a Galactic latitude of $b\sim13.5^\circ$, and so we can also study the period bimodality as a function of height above the Galactic mid-plane.
Both galaxy formation simulations \citep{ma2017} and observations of stars in the nearby Milky Way \citep{xiang2017} indicate that height above the mid-plane correlates strongly with the median ages for stars out to distances of several kpc. For low-mass stars the Milky Way thin disk has a scaleheight of $\sim$300 pc \citep{gilmore1983}.
However, since the \Kepler field is oriented towards low latitudes we do not reach significant {\it heights} above the disk.
The projected height above the mid-plane for the distance bins shown in Figures \ref{fig:color_period} and \ref{fig:per_hist} ranges from $Z\sim100$ pc at a distance of $d=350$ pc to $Z\sim230$ pc at $d=900$ pc. As a result, we are only sensitive to changes in the youngest stars within this span of $Z$.

In Figure \ref{fig:per_hist} (right) we find that the drop-off of the short-period (rapid rotating) component of the period bimodality is even more pronounced as a function of $Z$, decreasing rapidly after only 90 pc. With increasing height we also see that the shift to the longer period component is more smooth. However, from these two projections in Figure \ref{fig:per_hist} alone (distance and height) we cannot definitively determine the spatial structure of the rotation period bimodality.

\begin{figure}[!ht]
\centering
\includegraphics[width=3.4in]{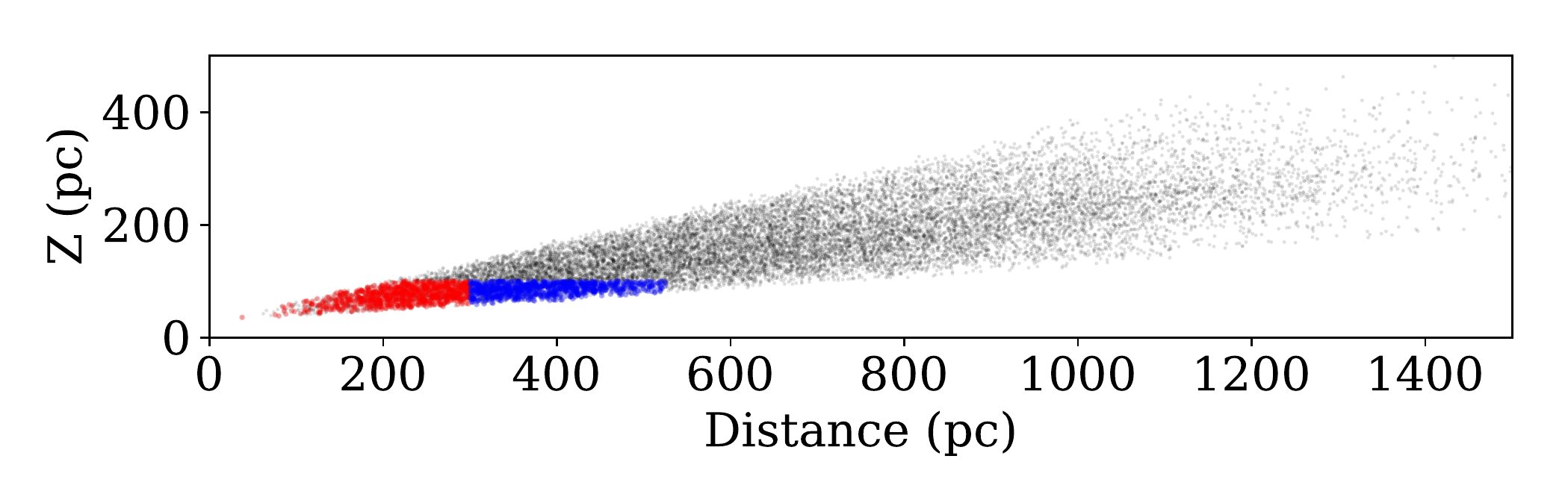}
\includegraphics[width=3.5in]{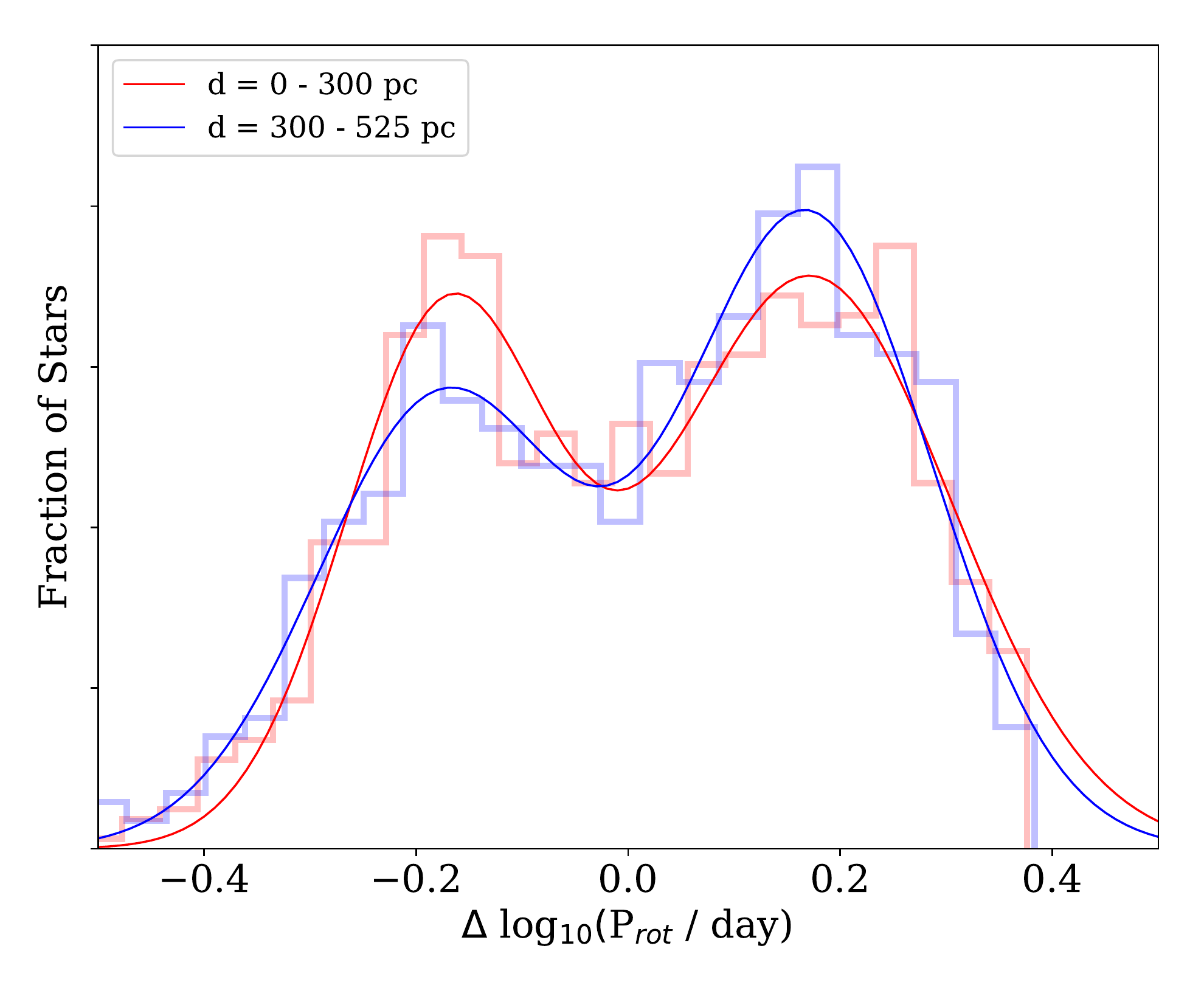}
\caption{
Top: Galactic height versus distance for the stars in our sample. Two approximately equal subsets of stars near the Galactic mid-plane have been highlighted, with distances of $d<300$ pc (red), and $300<d<525$ pc (blue).
Bottom: As in Figure \ref{fig:per_hist}, the distribution of rotation periods after a 600 Myr gyrochrone was subtracted (faint lines), and with two-Gaussian fits (bold lines) for the same subsets of stars as above. The period bidmodality is seen in both distance bins, suggesting it is not localized around the Sun.
}
\label{fig:dZ}
\end{figure}

To further understanding the spatial extent of this age-related feature, in Figure \ref{fig:dZ} we break the lowest height stars ($Z < 100$ pc) in to two roughly even samples, split as a function of their projected distance. Note we have also repeated this exercise for stars in higher ranges of $Z$, and find the decline of the rapid rotators is again uniform between subsamples of varying projected distance. The rotation period bimodality is clearly seen in both distance bins of Figure \ref{fig:dZ}, indicating that the feature is likely not a localized star formation history artifact centered around the Sun. Instead, we believe this feature is characteristic of the age--$Z$ dynamical correlation observed within the solar neighborhood of our Galaxy.

\section{Recalibrating Stellar Evolution Models}

\begin{figure*}
\centering
\includegraphics[width=7in]{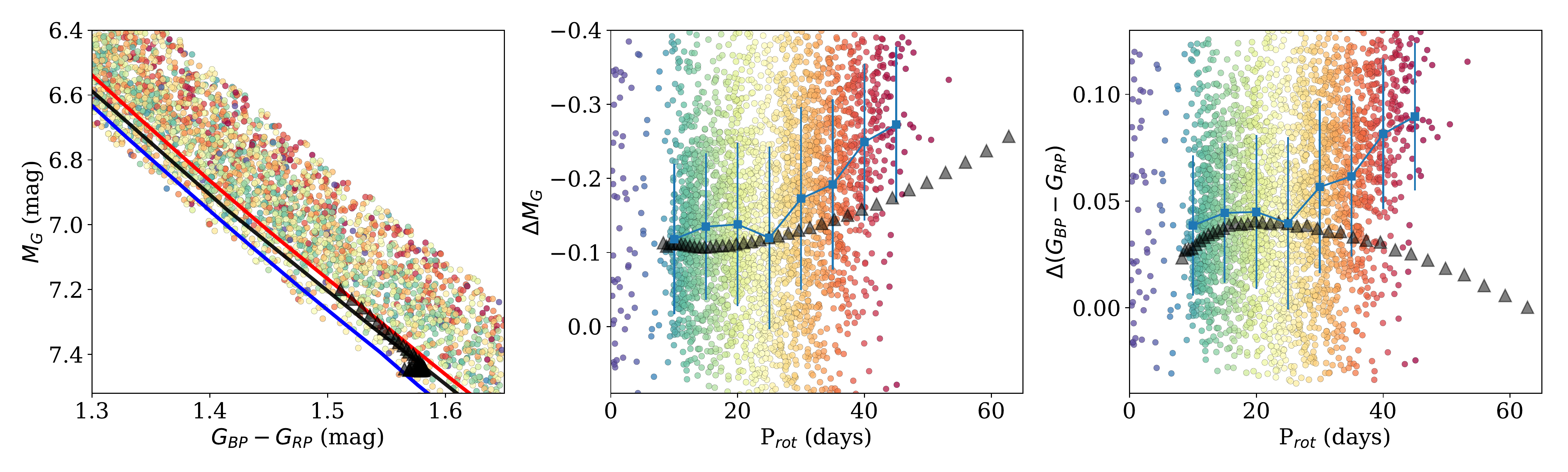}
\caption{
{\bf Left:} Enlarged portion of the color--magnitude diagram from Figure \ref{fig:cmd} in a region centered near $\sim$0.75 M$_\odot$, with main sequence stars colored by their measured rotation periods from \citet{mcquillan2014}.
MIST isochrones at ages of $10^8$, $10^9$, and $10^{10}$ yr are shown for comparison (blue, black, and red lines). The predicted evolution of a 0.7 M$_\odot$ star is highlighted (black triangles). 
{\bf Center:} Difference in $M_G$ from the $10^9$ yr MIST isochrone as a function of rotation period (point color again indicates rotation period, as in Left panel). An increase in the median brightness is seen in bins of rotation period (blue squares). Error bars are the standard deviation of $\Delta M_G$ in each bin. The 0.7 M$_\odot$ star brightness evolution from MIST is shown as a function of rotation evolution from \citet{meibom2009}, manually shifted to match the rapid rotating stars (black triangles).
{\bf Right:} Same as the center panel, but representing the change in $G_{BP}-G_{RP}$ color as a function of measured rotation period. Again the MIST--Meibom model evolution (black triangles) has been shifted to match the rapid rotators. Stars are predicted to get {\it bluer} at older ages, rather than redder as our sample indicates.
}
\label{fig:cmd_zoom}
\end{figure*}

A subtle feature we noticed in Figure \ref{fig:cmd} is the {\it diagonal} color gradient (i.e. rotation period gradient) for the stars in between the single and binary star main sequence populations. In Figure \ref{fig:cmd} this appears as a yellow stripe (i.e. rotation periods of 30-40 days) between these blue-green sequences for systems with colors of $G_{BP} - G_{RP} \approx 1.5$. To exaggerate this feature, we have reproduced a portion of our color--magnitude diagram focused on the main sequence near this stellar color in the left panel of Figure \ref{fig:cmd_zoom}. A clear color gradient is present, with red points (slower rotators) appearing preferentially above and to the right of the main sequence.
The center panel of Figure \ref{fig:cmd_zoom} demonstrates a correlation between the measured rotation period and the vertical offset (i.e. absolute magnitude) from the $10^9$ year MIST isochrone. Slower rotating stars are brighter at a given color.
However, as the right panel in Figure \ref{fig:cmd_zoom} shows, the MIST models predict a bluer color for stars at older ages, while our data show the slower rotators have redder colors on average.

In the center and right panels of Figure \ref{fig:cmd_zoom} we show a prediction of the brightness and color evolution of a 0.7 $M_\odot$ star as a function of its rotation period over time. The predicted evolution here is a combination of the MIST isochrone models from $10^8$ to $10^{10}$ years, as well as the \citet{meibom2009} gyrochronology model over this same time window. Note this  combined model has been arbitrarily offset in $\Delta M_G$ and $\Delta(G_{BP} - G_{RP})$ in the middle and right panels, respectively, to approximately match the CMD position observed for the rapid rotators. 

These data challenge our main sequence model evolution in two ways. Firstly, the diagonal gradient from the main sequence in rotation period is in conflict with the evolution predicted from MIST. The stellar evolution model shows that a star should evolve essentially {\it along} the main sequence track, as in the left panel of Figure \ref{fig:cmd_zoom}. The tension with our observed sample is highlighted in the right panel of Figure \ref{fig:cmd_zoom}, where the model predictions show stars becoming bluer in $G_{BP}-G_{RP}$ color as they age, while our slice through the observed rotation period sample suggests the older stars are instead redder.

We note that correlations in stellar rotation, age, and color have previously been reported for young clusters \citep[e.g.][]{stauffer2003,covey2016}.
However, these detections do not provide a straightforward explanation for the offset we detect in the {\em Gaia} CMD for field stars.
\citet{stauffer2003} detected an offset similar to the gradient we have observed for K dwarfs in the Pleiades and Praesepe open clusters, where younger, more rapidly rotating Pleiades members appear bluer/fainter in the ($B-V$, $V$) CMD than their older, more slowly rotating analogs in Praesepe.  
\citet{stauffer2003} suggest this offset reflects the influence of hot and cool star spots on a star's colors, as supported by the wavelength dependence of the offset. The color offset disappears in the ($V-I$, $V$) CMD and reverses in the ($V-K$, $V$) CMD, where the faster rotating Pleiades stars appear redder/brighter than their Praesepe counterparts.  
\citet{covey2016} extended this analysis to demonstrate that a Pleiades member's rotation period correlates with its relative position on the $V-K$ vs. $V$ cluster sequence, further supporting a picture in which rapidly rotating, magnetically active stars display different photospheric properties than their slower rotating, less magnetically active brethren.  The CMD gradient reported here are not easily explained as photospheric signatures of rotationally induced magnetic activity, however, because the gradients are most prominent among the most slowly rotating stars in the sample.  Indeed, we see no significant changes in the color or magnitude of stars with rotation periods shorter than $\sim$20 days; it is only among the more slowly rotating stars where color and luminosity gradients are significantly detected.  These slowly rotating stars presumably possess the lowest levels of magnetic activity, and thus have smaller, less prominent starspots to drive changes in the star's bulk photospheric properties.

This tension in the CMD evolution is compounded when we consider stellar metallicity, which we have not varied in our MIST model realization. Younger, presumably more metal rich stars should have redder optical colors on the main sequence, while older, more metal poor stars should be bluer. Instead we find that oldest, more slowly rotating stars are redder and brighter than their rapidly rotating counterparts. While we do not have independent metallicity constraints for the \Kepler stars in this sample, Figure \ref{fig:cmd_zoom} shows that age may be justa s important as metallicity in determining the precise CMD location for main main sequence stars. Likewise, stellar evolution models do not appear to reproduce the CMD positions for main sequence, low-mass stars of all ages.

The second, more subtle challenge presented by these data is the rotation periods observed for presumably older stars. 
We find that stars having $M_G$ offsets in the center panel of Figure \ref{fig:cmd_zoom} consistent with being several Gyr old have average rotation periods of 30--40 days, far shorter than the 60+ days values predicted by the spin-down model for a 0.7 $M_\odot$ star.  Note, \Kepler light curves are often not able to reliably measure rotation periods longer than $\sim$30 days. This bias means we may be missing even slower rotators from our \Kepler--{\em Gaia} combined sample, and do not know what the $M_G$ offset for such stars would be.

One possible interpretation of this result as observed here is that the older stars are spinning {\it faster} than expected for stars their age. This is qualitatively similar to the model of broken spin-down occurring at a critical Rossby number suggested by \citet{van-saders2016}.Though we cannot definitively confirm such intriguing rotation evolution from this initial investigation given the observation bias for long rotation periods from \Kepler, matching the {\em Gaia} CMD with other rotation period measurements may provide an ideal dataset to test the \citet{van-saders2016} model against.

\section{Discussion}

Using a sample of 16,248 single main sequence stars with measured rotation periods from \Kepler and parallaxes from {\em Gaia} DR2, we have begun to explore the spatial distribution of stellar ages near the Sun using gyrochronology. The bimodality in rotation periods first reported by \citet{mcquillan2013} appear to be constrained to low Galactic scaleheights, rather than in an obvious bubble centered around the Sun. Since height above the Galactic plane is assumed to be related to age, this is consistent with the rotation period bimodality being a direct tracer of the star formation history, and indicates a burst of star formation within the past $\sim$600 Myr within the Galactic disk.

Unfortunately the \Kepler survey only covers a single pointing, and so our volume analysis of the stellar ages is limited here. However, with 16 distinct lines of sight available from K2, the extended \Kepler mission, we will be able to more than double our sample of rotating field stars and probe a much wider range of stellar ages. This may enable us to determine the spatial scales over which star formation histories are coherant in the Milky Way disk. We could then make comparisons to the $\sim$100 pc resolved star formation history maps available for nearby galaxies such as Andromeda \citep[][]{lewis2015,williams2017}. For example, gyrochronology ages for disk stars within 1 kpc of the Sun could be used to test if star formation propagates due to spiral arm density waves \citep[e.g. see][]{choi2015}.

While we have removed as contaminants the prominent parallel main sequence of nearly equal-mass binary stars in this analysis, as well as sub-giant stars, both of these samples are clearly deserving of further analysis. For example, comparing the rotation period distribution for binary versus single main sequence stars may provide constraints on their dynamical histories. 

Finally, in the {\em Gaia} era, stellar evolution modelers have the daunting task of explaining an increasing number of precisely determined features in the CMD.
This includes the fascinating new main sequence gap revealed by \citet{jao2018}, as well as the diagonal age (or rotation) gradient in the main sequence shown in our Figures \ref{fig:cmd} and \ref{fig:cmd_zoom}.
Our work presents strong motivation for a new generation of stellar evolution and rotation models to accurately reproduce the main sequence as observed with \Kepler and {\em Gaia}. With \Kepler, K2, and soon TESS, precise rotation measurements will be available for hundreds of thousands of nearby stars that {\em Gaia} has provided reliable distances for. We hope these data will guide new isochrone models in reproducing both the observed colors and luminosities of stars due to their ages and abundances, as well as the evolution of their surface rotation rates.

\acknowledgments
The authors wish to thank Garrett Somers, Stephanie Douglas, Aaron Dotter, and Jennifer van Saders for their helpful suggestions and discussions during the preparation of this manuscript.

JRAD is supported by an NSF Astronomy and Astrophysics Postdoctoral Fellowship under award AST-1501418. 

This work made use of the \url{gaia-kepler.fun} crossmatch database, created by Megan Bedell.

This project was developed in part at the 2018 Gaia Sprint, hosted by the eScience and DIRAC Institutes at the University of Washington, Seattle, in parallel with the 2018 NYC Gaia Sprint, hosted by the Center for Computational Astrophysics at the Simons Foundation in New York City.

This work has made use of data from the European Space Agency (ESA) mission
{\it Gaia} (\url{https://www.cosmos.esa.int/gaia}), processed by the {\it Gaia}
Data Processing and Analysis Consortium (DPAC,
\url{https://www.cosmos.esa.int/web/gaia/dpac/consortium}). Funding for the DPAC
has been provided by national institutions, in particular the institutions
participating in the {\it Gaia} Multilateral Agreement.

\software{Python, IPython \citep{ipython}, NumPy \citep{numpy}, Matplotlib \citep{matplotlib}, SciPy \citep{scipy}, Pandas \citep{pandas}, Astropy \citep{astropy}}


\end{document}